\documentclass[%
 reprint,
nofootinbib,
 amsmath,amssymb,
 aps,
]{revtex4-1}

\usepackage{graphicx}
\usepackage{color}
\usepackage[colorlinks,citecolor=blue,urlcolor=blue]{hyperref}
\usepackage{nicefrac}
\usepackage{csquotes}
\MakeOuterQuote{"}
\graphicspath{{plots/}{./}}

\newcommand{\Neff}{N_{\rm eff}}
\newcommand{\ma}{m_{\rm a}}
\newcommand{\eV}{{\rm eV}}

\begin{document}

\title{New cosmological bounds on axions in the XENON1T window}

\author{Marius Millea}

\email{mariusmillea@gmail.com}
\affiliation{Berkeley Center for Cosmological Physics and Department of Physics, University of California, Berkeley, CA 94720, USA}

\date{\today}

\begin{abstract}
Motivated by a possible $\sim$\,eV-mass solar axion explanation to excess events recently detected  by the XENON1T experiment, I revisit and update cosmological constraints on axions in this mass range. I find that of the allowed XENON1T mass window (0.1\,--\,4.1\,eV for DFSZ axions and 46\,--\,56\,eV for KSVZ axions), only 0.1\,--\,0.35\,eV remains viable at 95\% confidence given current cosmological probes. If a 0.35\,eV DFSZ axion existed, it would be detectable at ${\sim}\,7\,\sigma$ via two independent physical effects with the next-generation CMB-S4 experiment. Conversely, even a combination of CMB-S4 with future DESI measurements falls just short of guaranteeing a 0.1\,eV-mass axion can be detected or ruled out. A future limit of $\Delta\Neff<0.027$ could rule out any generic axion-like particle across a wide range of masses as long as the reheating temperature is not too low, or alternatively, a future cosmological detection of such an axion-like particle could become the tightest existing observational lower bound on the reheating temperature. 
\end{abstract}

\maketitle

\section{Introduction}
Recently, the XENON1T experiment has reported a $3.5\,\sigma$ excess of electron recoil events in their detector over the expected background \cite{aprile2020}. One possible explanation is that these events come from hypothetical axions or axion-like particles (ALPs) produced in the Sun. The axion explanation, however, is in serious tension with constraints from stellar cooling \cite{giannotti2017}. Given this context, it is timely to review and update constraints on axions and ALPs from cosmology in the parameter range relevant for XENON1T. The constraints presented here are particularly valuable because they have an orthogonal set of systematics to both those from stellar cooling and from XENON1T, instead mainly depending only on very well understood linear perturbation theory, with only mild dependence on the underlying cosmological model.

Reference \cite{aprile2020} give axion-model-independent bounds on the coupling of ALPs to photons, electrons, and nucleons, as well as model-dependent fits to specific QCD axion models. The model-dependent fits yield a preferred mass-range of $46\,{<}\,m_a{/}\,\eV<56$ for KSVZ axions and $0.1\,{<}\,m_a/\eV\,{<}\,4.1\,\rm eV$ for DFSZ axions (these specific models are discussed in the next section). Cosmological bounds on QCD axions in this mass range have also been considered by \cite{chang1993,hannestad2007,melchiorri2007,hannestad2008,hannestad2010,archidiacono2013,giusarma2014,divalentino2016b}. This work builds upon these, updating, clarifying, and presenting some new ways to view and understand the cosmological axion bounds. I will also consider generic ALP constraints, where cosmological probes can place unique model-independent bounds. 

\section{Background}

The axion is a hypothetical particle which arises in the Peccei-Quinn solution to the strong CP problem in quantum chromodynamics (QCD) \cite{peccei1977,weinberg1978,wilczek1978,peccei2008}. As the pseudo Nambu-Goldstone boson of a new global $\rm U(1)_{PQ}$ symmetry which is spontaneously broken, the axion acquires a mass given by
\begin{align}
    m_{a}=\frac{f_{\pi} m_{\pi}}{f_{a}} \frac{\sqrt{z}}{1+z},
\end{align}
where $f_a$ is the axion decay constant, $f_\pi\,{=}\,92\,\rm MeV$ is the pion decay constant, $m_\pi\,{=}\,135\,\rm MeV$ is the pion mass, and $z=m_u/m_d\simeq 0.56$ is the ratio of up to down quark masses. Axions generically couple to standard-model particles, with some model-dependence in the form of the couplings, how PQ charges are distributed to the standard model (SM) particles, and whether any other new symmetries are introduced. These coupling are such that remarkably complimentary constraints exist between 1) axions produced in the Sun and subsequently detected by experiments such as XENON1T and 2) axions produced in the first few minutes after the big bang and subsequently detected via their imprint on cosmological observables. 

\begin{figure}
    \includegraphics[width=\columnwidth]{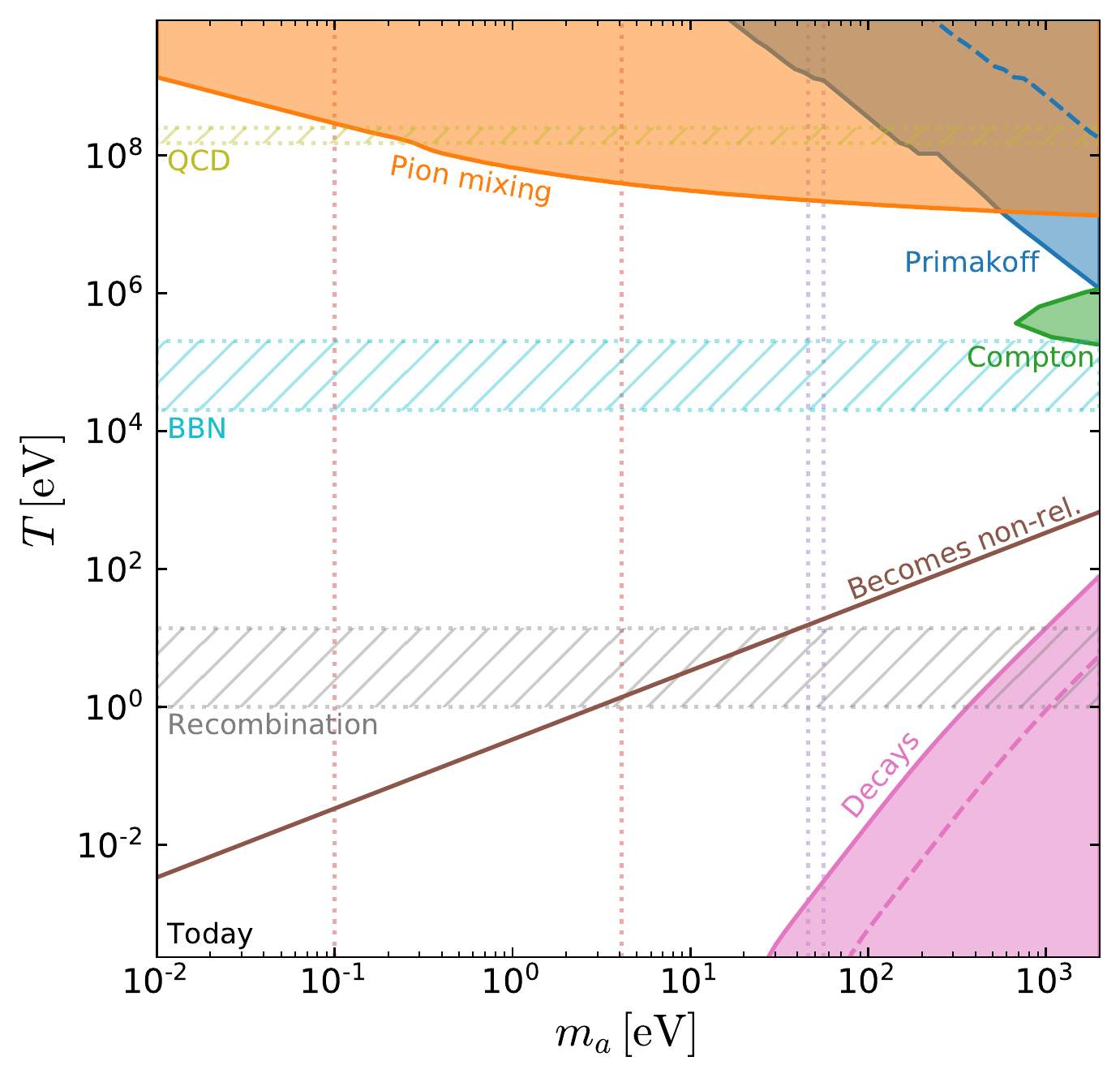}
    \caption{Processes keeping QCD axions in equilibrium and important epochs in the QCD axion evolution. Shaded regions indicate where the labeled process is active. In the case that these depend on the model-dependent anomaly factors, the solid line is for the baseline DFSZ case with $E/N\,{=}\,8/3$, and the dashed line is for the baseline KSVZ case with $E/N\,{=}\,2$. Hatched bands denote the approximate extent of the QCD phase transition, BBN, and CMB recombination, and the very bottom of the plot corresponds to the temperature today. The left and right pairs of dotted vertical lines enclose the mass region preferred by XENON1T for DFSZ and KSZV axions, respectively.}
    \label{fig:ma_Tfo}
\end{figure}

The most relevant couplings in terms cosmological constraints are the couplings to photons, electrons, and pions. The photon coupling gives rise to three processes which should be considered: 1) axion production via the Primakoff effect, wherein photons are converted to axions in the presence of charged particles, $q\gamma\,{\rightarrow}\,q a$, 2) axion production via inverse decays, $\gamma\gamma\,{\rightarrow}\,a$, and 3) axion decay to photons, $a\,{\rightarrow}\,\gamma\gamma$,

Inverse decays turn out to be unimportant for masses $m_a\lesssim\rm keV$ \cite{cadamuro2012,millea2015}. Conversely, the per-particle rate for forward decays can be calculated as
\begin{align}
    \label{eq:decay}
    \Gamma_{a\rightarrow\gamma\gamma} = \frac{m_a^3 g_{a\gamma}^2}{64 \pi},
\end{align}
where the photon coupling coefficient is
\begin{align}
    \label{eq:gag}
    g_{a \gamma}&=\frac{\alpha}{2 \pi f_{a}}\left(\frac{E}{N}-\frac{2}{3} \frac{4+z}{1+z}\right).
\end{align}
and $E$ and $N$ are model-dependent electromagnetic and color anomaly factors. Two typically considered models are the KSVZ \cite{kim1979,shifman1980} and DFSZ \cite{zhitnitskij1980,dine1981} models. In terms the XENON1T data, the DFSZ fit assumes a standard value of $E/N\,{=}\,8/3$ and the KSVZ fit requires $E/N\,{=}\,2$, the latter which is a special case since a coincidental cancellation in \eqref{eq:gag} gives particularly weak photon coupling \cite{cheng1995,diluzio2017}. For the remainder of the work, I will assume these two values for each case, respectively. 

Decays become effective when they are more rapid than the Hubble rate, which happens at a temperature $T_{\rm fo}$ defined by
\begin{align}
    \label{eq:freezeout}
    \Gamma(T_{\rm fo}) = H(T_{\rm fo}).
\end{align}
Fig.~\ref{fig:ma_Tfo} shows numerical solutions of \eqref{eq:freezeout}, such that vertical slices can be used to read off the thermal history for axions of different masses. Shaded regions indicate temperatures where different interactions are effective, and in the case that these regions are model-dependent, solid lines are for the DFSZ axion and dashed lines are for the KSVZ axion. For the photon decays of present interest, this demonstrates that axions are still stable today for $m_a\lesssim \rm 80\,eV$ in the KSVZ case, and $m_a\lesssim \rm 30\,eV$ in the DFSZ case. The preferred XENON1T masses in both scenarios are small enough to obey these limits, thus no late-time decays need to be considered. For even higher masses, decays can happen during or well before recombination, giving rise to a rich phenomenology depending on the exact decay epoch and whether the axions are still relativistic at this point. These models are explored in \cite{cadamuro2011a,cadamuro2012,millea2015,depta2020}.

This leaves the Primakoff effect as the last photon process to consider. The scattering rate for Primakoff production has been estimated by \cite{bolz2008,cadamuro2011a} to be
\begin{align}
    \label{eq:primakoff}
    \Gamma_{q\gamma\rightarrow qa} &\simeq \frac{\alpha g_{a\gamma}^{2} \pi^{2}}{36 \zeta(3)}\left(\log \left(\frac{T^{2}}{m_{\gamma}^{2}}\right)+0.82\right) n_{q},
\end{align}
where $n_q\,{=}\,\Sigma_i Q^2_i n_i$ is the number density of all charged particle species weighed by their squared charge, and $m_\gamma\,{=}\,T/(6\alpha\sqrt{g_q(T)})$ is the plasmon mass of the photon, where $g_q(T)$ is the effective number of charged degrees of freedom in the plasma, defined such that $n_q\,{=}\,\zeta(3) g_q(T) T^3/\pi^2$. Fig.~\ref{fig:ma_Tfo} shows that there is almost no window where Primakoff production is the dominant production mechanism for QCD axions (at least not for the values of $E/N$ assumed here), but it will be more relevant for generic ALPs in Sec.~\ref{sec:alps}.

The more important coupling for QCD axions, especially in the mass range relevant for XENON1T, is the axion-pion coupling. This arises due to the axion coupling to gluons, which will necessarily be present for axions that solve the strong CP problem. The processes to consider are $\pi^0\pi^\pm\rightarrow a \pi^\pm$ and $\pi^+\pi^-\rightarrow a \pi^0$, with an approximate total scattering rate for both given by \cite{chang1993,hannestad2005}:
\begin{multline}
    \Gamma_{\pi\pi\rightarrow a\pi} = \frac{3}{1024 \pi^{5}} \frac{1}{f_{a}^{2} f_{\pi}^{2}} C_{a \pi}^{2} n_{a}^{-1} T^{8}  \times \\
    \int_0^\infty d x_{1} d x_{2} \frac{x_{1}^{2} x_{2}^{2}}{y_{1} y_{2}} f(y_{1}) f(y_{2})  \int_{-1}^{1} d \omega \frac{\left(s-\mu_{\pi}^{2}\right)^{3}\left(5 s-2 \mu_{\pi}^{2}\right)}{s^{2}}
\end{multline}
where $\mu_\pi=m_\pi/T$, $s=2\left(\mu_{\pi}^{2}+\left(y_{1} y_{2}-x_{1} x_{2} \omega\right)\right)$, and $C_{a\pi}\,{=}\,(1-z)/(3(1+z))$. The freeze-out for this process is also shown in Fig.~\ref{fig:ma_Tfo}, and turns out to be the most important process for setting the relic abundance in the mass range of interest. A nice consequence of constraints being driven by the pion coupling is that this coupling does not depend on the anomaly factors, and is thus more generic.

Finally, Fig.~\ref{fig:ma_Tfo} shows the temperature at which axions with different masses become non-relativistic. With this information, the overall picture for the axions in the XENON1T mass range is thus set: they decouple shortly after the QCD phase transition, they become non-relativistic during or somewhat after cosmic microwave background (CMB) recombination, and they remain stable until today.

\section{Results for QCD axions}

\begin{figure}[ht!]
    \includegraphics[width=\columnwidth]{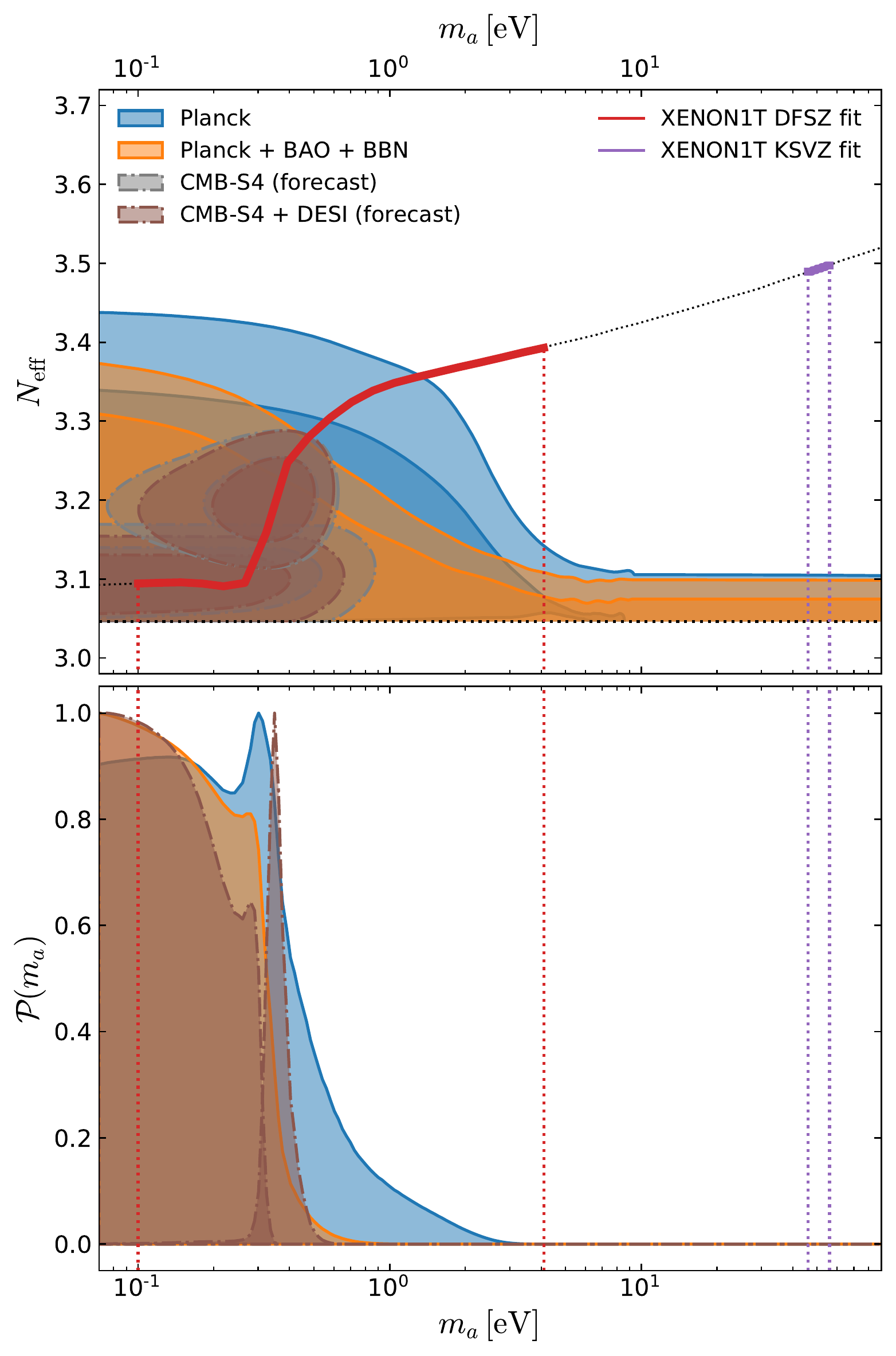}
    \caption{({\it Top panel}) Joint 1 and 2\,$\sigma$ confidence contours on the ALP mass, $m_a$, and the effective number of relativistic species, $\Neff$ (including the ALP and standard model neutrinos). QCD axions lie along the black dotted line in this parameter space, and the range of masses favored by XENON1T for KSVZ and DFSZ axions are shown as the pairs of purple and red vertical lines, respectively. Large parts of these mass ranges are ruled out by {\it Planck} alone (blue contours) or in combination with BOSS BAO and primordial BBN abundance measurements (orange contours). In the future, CMB-S4 and DESI measurements could make a ${\sim}\,7\sigma$ detection of the DFSZ axion if its mass is on the upper end of the mass range preferred by XENON1T (brown dot-dashed contours centered on 0.35\,eV), or further limit the allowed mass if it is on the lower end (brown dot-dashed contours centered on 0.1\,eV). ({\it Bottom panel}) Constraints on the mass of the QCD axion. Note that these are slices through the top panel along the black dotted line, rather than a marginalization over $\Neff$.}
    \label{fig:Neff_ma}
\end{figure}

The phenomenology just described is all that is necessary to compute the cosmological impact of axions, which turns out to be fairly simple. Because in all cases they are decoupled and relativistic through big bang nucleosynthesis (BBN), their impact on BBN is exactly captured by the standard $\Neff$ parameter, which controls the energy density in all relativistic species relative to photons,
\begin{align}
    \rho_{\rm rad} = \rho_\gamma \left[1+\Neff\frac{7}{8}\left(\frac{4}{11}\right)^{4/3}\right]
\end{align}
The three SM neutrino species contribute $\Neff\,{=}\,3.046$. The axion contribution is, by definition, 
\begin{align}
    \label{eq:DeltaNeff}
    \Delta \Neff \frac{7}{8}\left(\frac{4}{11}\right)^{4/3} &\equiv \frac{\rho_a}{\rho_\gamma} = \frac{1}{2} \left(\frac{T_a}{T_\gamma}\right)^4 \\ &= \frac{1}{2} \left(\frac{47/11}{g_\star(T_{\rm fo})}\right)^{4/3}
\end{align}
where $g_\star(T_{\rm fo})$ is the number of degrees of freedom in the particle species still present and relativistic in the primordial plasma when the axion freezes out. The last equality follows because the subsequent heating of the plasma by the annihilations of these degrees of freedom happens in equilibrium and hence conserves comoving entropy density  \cite[e.g.][]{abazajian2016}. Computing $T_{\rm fo}$ as a function of the axion mass as in Fig.~\ref{fig:ma_Tfo} gives
\begin{alignat}{3}
    0.048\,{<}\,&\Delta \Neff\,{<}\,0.35 {\rm \;\;for\;\;} &0.1\,{<}\,& m_a/\eV\,{<}\, 4.1 {\rm \;\;and} \\
    0.44\,{<}\,&\Delta \Neff\,{<}\,0.45 {\rm \;\;for\;\;} &46\,{<}\,& m_a/\eV\,{<}\,56.
\end{alignat}
The full dependence of $\Neff$ on $m_a$ is shown in Fig.~\ref{fig:Neff_ma} in dotted-black, amounting to what can be considered a "QCD axion consistency" relation. Note the drop near $m_a\,{\sim}\,0.4\,\rm eV$, which corresponds the axion-pion freeze-out temperature coinciding with the drop in $g_\star$ at the end of the QCD phase transition.

A recent BBN-only bound on $\Neff$ is given by \cite{fields2020}, who combine primordial helium and deuterium abundance measurements, marginalizing over the baryon-to-photon ratio. This yields $\Neff\,{=}\,2.878\,{\pm}\,0.278$, which then translates to an upper limit, 
\begin{align}
    \label{bounds:bbn}
    m_a < 12.7\,\rm eV \;\; (BBN; \; 95\%)
\end{align}
This rules out the entire XENON1T KSVZ mass range at 95\%, although allows it at 99\%. 

The BBN constraint is particularly important because it assumes nothing about the late-time behavior of these axions. Although for the values of $E/N$ assumed here they are stable until today and behave like hot dark matter, for other reasonable values they might decay sometime between recombination and today, their decay photons constituting an extragalactic background light or contributing monochromatic lines to spectra of astrophysical objects. Searches for these imprints with varying assumptions on the axion couplings have excluded some, but not all, of the mass range in question \cite{turner1987,ressell1991,bershady1991,overduin1993,grin2007,cadamuro2012}. Modeling these late-time decays could, at least in theory, invalidate some of the CMB and BAO bounds discussed below, but the BBN bound is insensitive to this and creates a unified ruled-out cosmological mass bound without the need for such modeling.

\paragraph*{With CMB and other datasets} The bounds in \eqref{bounds:bbn} can be further tightened up by including other lower-redshift data. Here, I will consider various combinations of:
\begin{itemize}
    \item \textsc{Planck} -- \textit{Planck} 2018 TT, TE, EE power spectra as well as the reconstructed gravitational lensing potential power spectrum \cite{planckcollaboration2018}, 
    \item \textsc{BAO} -- The baryon acoustic oscillation sample used in \cite{planckcollaboration2018}, consisting of BOSS DR12 anisotropic BAO measurements \cite{alam2017} and angle-averaged quantities from 6dFGS and SDSS-MGS \cite{beutler2011,ross2015}.
    \item \textsc{SH$_0$ES} -- A prior of $73.48\,{\pm}\,1.66 \; \rm km/s/Mpc$ \cite{riess2018}
\end{itemize}
Although axions in the XENON1T window remain stable until today, they do become non-relativistic, so for the purpose of CMB and other late-time cosmological bounds, they cannot simply be modeled as extra radiation as in the BBN case. While a fully correct treatment would include axion density perturbations in the set of equations solved by typical Boltzmann codes to compute CMB and matter power spectra, there is a simpler approach. Axions turn out to be, to a very good approximation for current and near-future precision measurements, equivalent to massive sterile neutrinos. This allows reusing existing code and even already computed constraints. 

A typically used parametrization for massive sterile neutrinos is $(\Delta\Neff, m_{\rm sterile}^{\rm eff})$ where $m_{\rm sterile}^{\rm eff}$ controls the present-day energy density in sterile neutrinos \cite{planckcollaboration2018}, 
\begin{align}
    \label{eq:meffsterile}
    \Omega_\nu h^2 \equiv \frac{m_{\rm sterile}^{\rm eff}}{94.1\,\eV}
\end{align}
By comparison, the present-day energy density in a thermalized non-relativistic bosonic species, $a$, with one degree of freedom, which decoupled while relativistic (i.e., an axion), can be written as 
\begin{align}
    \Omega_a h^2 = m_a n_a &\simeq  \left(\frac{m_a}{94.1\,\eV} \right) \left(\frac{11}{6}\right) \left(\frac{T_a}{T_\gamma}\right)^3 \\
    \label{eq:omega_a}
    &\simeq 1.014 \left( \frac{m_a}{94.1\,\eV} \right) \Delta\Neff^{\,3/4}
\end{align}
where the last equality follows from \eqref{eq:DeltaNeff}. Equating \eqref{eq:meffsterile} with \eqref{eq:omega_a} implies that existing constraints on $m_{\rm sterile}^{\rm eff}$ can be mapped on to constrains on $m_a$ via
\begin{align}
    \label{eq:ma_reparam}
    m_a = m_{\rm sterile}^{\rm eff} \Delta\Neff^{\,-3/4} / 1.014
\end{align}
A caveat to this is that sterile neutrinos have a Fermi-Dirac phase space distribution, while axions instead obey Bose-Einstein statistics. However, as demonstrated by \cite{desalas2018}, state-of-the-art CMB data cannot currently distinguish whether the energy density in even all three SM neutrinos is Fermi-Dirac or Bose-Einstein, so it certainly is insensitive to the statistics of just the small fraction of the energy density contained in $\Delta\Neff$. A similar result for matter clustering is shown by \cite{hannestad2005a}, the only exception being in the non-linear regime of halo cores, which are not considered here. Thus, it is safe to reuse sterile neutrino code and constraints for the purpose of constraining axions here.

I begin with the public \textsc{Planck} Monte-Carlo Markov chains assuming the $\Lambda$CDM\,{+}\,$\Neff\,{+}\,m_{\rm eff}^{\rm sterile}$ model. Constraints after the reparametrization to $(\Neff,m_a)$ are shown in Fig.~\ref{fig:Neff_ma}. While previous analyses constrained $m_a$ directly, it is useful to keep the two parameters separate to better understand the physical origin of constraints. CMB constraints on $\Neff$ arise from changes to the expansion rate near recombination and subsequent impacts to CMB diffusion damping \cite{hou2013}, as well as a phase shift of the acoustic peaks due to supersonic propagation of spatial perturbations in the axion density field \cite{follin2015,baumann2016b}. The CMB sensitivity to the axion mass arises because at late times, the axions act as hot dark matter, changing the expansion rate as they become non-relativistic and suppressing structure up to their free-streaming scale. This effect is particularly expressed in the gravitational lensing of the CMB \cite{pan2015}. The degeneracies between the two parameters in Fig.~\ref{fig:Neff_ma} represents the fact that if $\Delta \Neff$ is small, there is very little energy density in the relic axions at all, making it harder measure their mass. 

Direct constraints on $\ma$ for QCD axions can be obtained by computing the probability along a slice through the joint $(\Neff, m_a)$ space following the QCD axion consistency line. This is shown in the bottom panel, giving upper bounds of
\begin{alignat}{2}
    \label{bounds:planckbbn} 
    \ma &< 0.8\,\rm eV \;\;  &&(\text{\small \textsc{Planck}; 95\%}) \\
    \label{bounds:planckbbnbao}
    \ma &< 0.37\,\rm eV \;\; &&(\text{\small \textsc{Planck}+BBN+BAO; 95\%}) 
\end{alignat}
The bounds in \eqref{bounds:planckbbnbao} are the tightest existing cosmological constraints on $\ma$, tighter than previous ones using similar datasets \cite{hannestad2010,divalentino2016b,giusarma2014} mainly due to the inclusion of more recent {\it Planck} data. They rule out the KSVZ window at extremely high significance, and most (but not all) of the DFSZ window at ${>}\,5\,\sigma$ as well.

\paragraph*{Neutrino masses} The bounds in \eqref{bounds:planckbbn} and \eqref{bounds:planckbbnbao} assume neutrinos have the normal hierarchy and the minimum possible masses. Because cosmological bounds are mainly sensitive to the total energy density in neutrinos plus axions, and because the neutrino masses could only be {\it higher} than the minimum, marginalizing over the sum of the masses of the active neutrinos, $\Sigma m_\nu$, can only {\it improve} constraints on $m_a$. Using \textsc{CosmoMC} \cite{lewis2002} to run a new chain assuming the $\Lambda$CDM\,{+}\,$\Neff\,{+}\,m_{\rm eff}^{\rm sterile}\,{+}\,\Sigma m_\nu$ model (a case not already available), I find that, with a flat prior on $\Sigma m_\nu$,
\begin{align}
    \label{bounds:planckbbnbao_mnu}
    \ma &< 0.35\,\rm eV \;\; (\text{\small {\rm \textsc{Planck}+BBN+BAO; $\Sigma m_\nu$; 95\%}})
\end{align}
Note, however, that the level of improvement is fairly sensitive to the prior on the neutrino mass because the data poorly constrains $\Sigma m_\nu$ alone, leaving the constraint more prior-driven.

\paragraph*{Hubble tension} The presence of the axion does not appear to have a big impact in resolving possible tensions in the Hubble constant \cite[for a summary of the tension, see e.g.][]{knox2020}. The mean value of $H_0$ is not appreciably increased from its $\Lambda$CDM value in any of cases discussed thus far, indicating that the $(\Neff,m_a)$ extension would have poor Bayesian evidence relative to $\Lambda$CDM when adding \textsc{SH$_0$ES} data. Combining the two datasets anyway should shift $\Neff$ to higher values as per conventional wisdom \cite{riess2016}, as well as tightening the $m_a$ constraint since more massive relics end up reducing the late time expansion rate \cite{planckcollaboration2018}. The former suggests a loosening of the QCD axion mass bound while the latter the opposite. The interaction of these shifts with the shape of the QCD axion consistency line ultimately gives overall slightly tighter constraints, reducing the upper bound on $m_a$ from 0.37\,eV to 0.36\,eV.

\paragraph*{Forecasts}
Can cosmological bounds do better in the future? The next-generation CMB-S4 experiment aims to measure the CMB temperature and polarization across ${\sim}\,50\%$ of the sky to noise levels almost two orders of magnitude better than {\it Planck} and to angular scales more than twice as small \cite{abazajian2019a}. In terms of these axion constraints, the slope of the QCD axion consistency line at $\ma\,{<}\,0.4\,\rm eV$ in Fig.~\ref{fig:Neff_ma} indicates that improvements can come both from better measurements of the early-time axion contribution to $\Neff$, as well as the late-time impact of its mass. CMB-S4 will be more sensitive to both of these effects \cite{baumann2016b,pan2015}. The next-generation spectroscopy survey DESI \cite{desicollaboration2016} is currently under way as well, and will also be sensitive to the presence of relic axions due to their impact on late-time expansion and structure growth. 

To begin, the gray contours in Fig.~\ref{fig:Neff_ma} show the results of a Fisher forecast for CMB-S4\footnotemark[3]. Two sets of contours\footnotemark[4]\ are shown for two different choices of fiducial $m_a$. The level of constraint depends significantly on this choice since if $m_a$ is small, there is less signature of axions to detect at all. For a fiducial $\ma\,{=}\,0.35\,\rm eV$ right on the upper bound of the currently allowed DFSZ window, CMB-S4 alone would detect the presence of the axion at ${\sim}\,7\,\sigma$ with a $\sigma(m_a)\,{=}\,0.055\,\eV$. Both non-zero $\Delta\Neff$ and $m_a$ would be detected independently at ${\sim}\,4\,\sigma$, which would offer a powerful consistency check giving confidence that the detection was indeed of a QCD axion. In this case, the addition of a DESI forecast\footnotemark[5]\ only marginally helps improve the constraint on $m_a$. On the other hand, if the true axion mass is at the lower end of the preferred window, $m_a\,{=}\,0.1\,\eV$, then no detection can be guaranteed. DESI in this case does help somewhat, but the fact that the QCD axion consistency line turns up exactly in this region means improvements are not as large as they could be otherwise; the 95\% upper limit reduces from 0.3 to 0.25\,\eV.

\footnotetext[3]{The CMB-S4 forecast uses temperature (T), polarization (P), and lensing reconstruction power spectra, assuming 1\,$\mu$K-arcmin T noise ($\sqrt{2}$ higher for P), 2$^\prime$ beams, $f_{\rm sky}\,{=}\,0.5$, and $50\,{<}\,\ell\,{<}\,3000$ for T and $50\,{<}\,\ell\,{<}\,5000$ for P. This is complemented with {\it Planck} on the remaining $f_{\rm sky}=0.25$ and on the full sky at $\ell\,{<}\,50$.}
\footnotetext[4]{In such a degenerate space, one should not necessarily expect that actual future posteriors look exactly like this, but should instead keep in mind the precise statistical meaning of the Fisher information, mainly that it gives the lower bound on the variance of an unbiased estimator for $\Neff$ and $\ma$.}
\footnotetext[5]{The DESI forecast uses the errors on the transverse and radial BAO distance scales from galaxies, Lyman-$\alpha$ forest, and high-$z$ quasars described in \cite{font-ribera2014} in conjunction with constraints from redshift-space distortions following the procedure from \cite{pan2015}.}

\section{Results for axion-like particles}
\label{sec:alps}

The previous section assumed the particles in question were QCD axions, whose couplings are all controlled (up to some model-dependent factors) by a single parameter, the axion decay constant, $f_a$. In this case, it was necessarily the axion-pion coupling which set the relic abundance and dictated the achievable level of constraints. ALPs are a more general class of particles which can arise in many theories with spontaneously broken symmetries \cite[for an overview, see e.g.][]{ringwald2014,baumann2016b,marsh2017}, and where the couplings need not be related in the same way. 

For simplicity, I consider only the ALP photon coupling here, but interesting bounds exist based on other couplings as well (e.g., contemporaneous results of \cite{arias-aragon2020}). The photon coupling leads to ALP production via the Primakoff effect, with a crucial feature that the freeze-out temperature is independent of the ALP mass \eqref{eq:primakoff}. As before, CMB constraints on $\Delta\Neff$ bound the freeze-out temperature, which in turn is a (unique) direct bound on just the photon coupling, $g_{a\gamma}$. Current measurements suggest $\Delta\Neff\,{<}\,0.44$ at 95\% confidence, which translates to
\begin{align}
    \label{bounds:gag}
    g_{a\gamma} &< 6.6\times10^{-8}\,\rm GeV^{-1} \;\; (\text{\small {\rm \textsc{Planck}+BBN+BAO; 95\%}})
\end{align}
Two caveats should be noted here. First, although the production rate is independent of mass, the decay rate to photons is not \eqref{eq:decay}. To ensure decays do not confuse the late-time interpretation of the cosmological data in some way, this bound can be viewed as only valid when $m_a\,{<}\,4\,\eV$, which is the necessary requirement for the ALPs to remain stable. Second, it should be highlighted (for this and the previous section as well) that the production of these ALPs requires the universe to have been hot enough, thus the reheating temperature cannot be below the freeze-out temperature for the $g_{a\gamma}$ in \eqref{bounds:gag}. This places another requirement, mainly, $T_{\rm R}\,{>}\,20\,\rm MeV$. Although more difficult to visualize here, a nice alternative to interpreting the conditionality of these bounds is to jointly constrain $g_{a\gamma}$ and $T_{\rm R}$ as in \cite{grin2008}.

\begin{figure}
    \vspace{1cm}
    \includegraphics[width=\columnwidth]{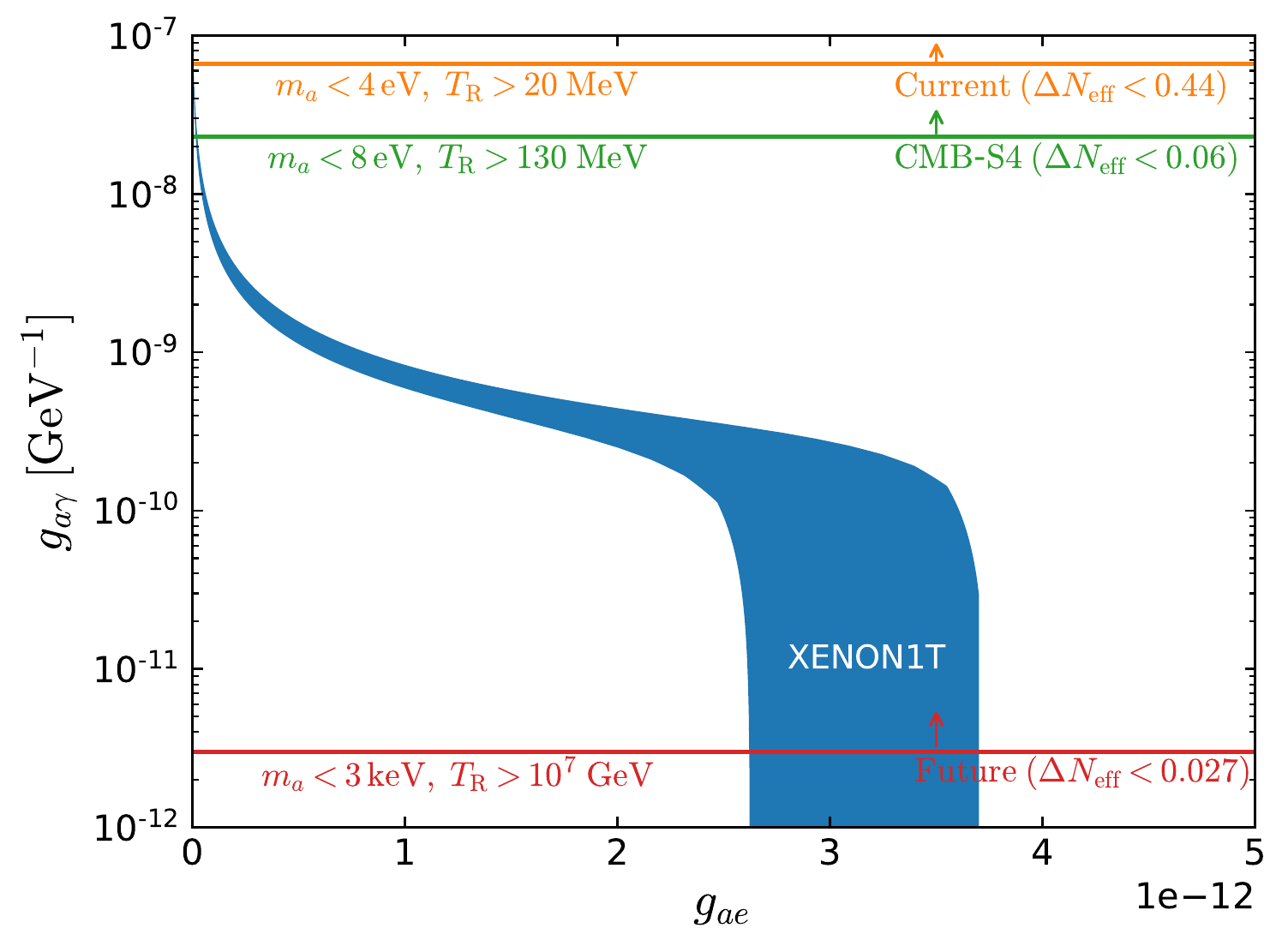}
    \caption{Cosmological bounds on generic ALPs as compared to the XENON1T preferred region. Each bound only applies if the specified requirement on mass, $m_a$, and reheating temperature, $T_{\rm R}$, is met. Arrows indicate the 95\% excluded region. The ``Future'' contour can be pushed arbitrarily further down if the reheating temperature is higher; the choice here of $T_{\rm R}\,{=}\,10^7$ is only to have a reference within this plot range.}
    \label{fig:gae_gag}
\end{figure}

Fig.~\ref{fig:gae_gag} shows the bounds from \eqref{bounds:gag} relative to the XENON1T preferred region, demonstrating that present generic ALP constraints are not at a level to be informative. This figure also shows the forecast for CMB-S4, which pushes slightly into the XENON1T region and also relaxes the mass and reheating temperature requirements. Ultimately, with a bound that excludes $\Delta\Neff\,{=}\,0.027$ (the minimum possible value for any relic which was at some point in thermal equilibrium with the rest of the standard model) one could rule out the entire XENON1T region to effectively arbitrarily low values of $g_{a\gamma}$, modulo the aforementioned assumptions about reheating, as well as assuming no other non-SM particles decayed after axion freezeout. Although a challenging measurement, this highlights the conclusions of \cite{baumann2016a,abazajian2016} about the wide-ranging impact such a measurement could have.

\section{Conclusion}

In this work, I have discussed cosmological constraints on axions and axion-like particles in the parameter space relevant for the XENON1T experiment. Only the 0.1\,{--}\,0.35\,eV range remains viable cosmologically for a possible QCD axion explanation of the XENON1T results. This currently-tightest such bound marks another step in the steady march of increasingly precise cosmological constraints on the axion mass. I have demonstrated that it is both practically simple and physically useful to map axion parameters onto typical sterile neutrino parameters $(\Neff, m_{\rm sterile}^{\rm eff})$. This allows one to view constraints in terms of a QCD axion consistency relation, which also makes clearer how constraints can improve in the future: in the near term, better bounds on the early-time relativistic energy density $\Neff$ can still help, however, unless the threshold $\Delta\Neff\,{<}\,0.027$ can be broken, the main longer-term improvements will come from measuring the late-time impact of the axion mass. If the axion exists and its mass is on the upper end of the currently allowed region, CMB-S4 will make a high significance detection. However, the low end of the mass region cannot be guaranteed to be detected or excluded. Although the forecasts performed here focused on simple and robust cosmological probes, they by no means represent an exhaustive list of planned future measurements. Given the focus of future measurements on the mass of the neutrinos and given that the axion mass is measured through similar physical effects, it seems plausible that with more lower-redshift clustering data and a more aggressive push into non-linear regimes than considered here, an upper bound below 0.1\,eV could be achieved.

\acknowledgements
I thank Uros Seljak, Bradley Kavanagh, and Brian Fields for discussions, encouragement, and comments on a draft, which greatly helped this work.

\bibliography{xenonalps}

\end{document}